\documentclass[12pt]{article}

\usepackage{amssymb,amsmath}

\usepackage{graphicx}
\usepackage{url}

\usepackage{cite}

\newcommand{\be}{\begin{equation}}
\newcommand{\ee}{\end{equation}}

\newcommand{\dlt}{\delta}
\newcommand{\prt}{\partial}
\newcommand{\br}{{\bf r}}

\newcommand{\bt}{\beta}
\newcommand{\vp}{\varphi}
\newcommand{\ep}{\varepsilon}
\newcommand{\al}{\alpha}
\newcommand{\ra}{\rightarrow}
\newcommand{\sgm}{\sigma}

\newcommand{\om}{\omega}

\newcommand{\cH}{{\cal H}}

\newcommand{\lgl}{\langle}
\newcommand{\rgl}{\rangle}

\begin{document}

\begin{center}

{\Large{\bf Mode interference in quantum joint probabilities\\ for multimode
Bose-condensed systems} \\ [5mm]

V.I. Yukalov$^{1,2}$, E.P. Yukalova$^{1,3}$ and D. Sornette$^{1,4}$} \\ [3mm]

{\it
$^1$D-MTEC, ETH Z\"urich, Swiss Federal Institute of Technology, \\
Z\"urich CH-8092, Switzerland \\ [3mm]

$^2$Bogolubov Laboratory of Theoretical Physics, \\
Joint Institute for Nuclear Research, Dubna 141980, Russia \\ [3mm]

$^3$Laboratory of Information Technologies, \\
Joint Institute for Nuclear Research, Dubna 141980, Russia \\ [3mm]

$^4$Swiss Finance Institute, c/o University of Geneva, \\
40 blvd. Du Pont d'Arve, CH 1211 Geneva 4, Switzerland}

\end{center}

\vskip 1cm

\begin{abstract}

The method of defining quantum joint probabilities of two events is applied to
a multimode system of trapped Bose-condensed atoms. The coherent modes are
generated by modulating the trapping potential with an alternating field with
a frequency in resonance with a transition frequency between two coherent modes.
The system is subjected to nondestructive measurements randomly influencing the
phase difference of the modes. The joint probability of observing two events,
corresponding to the observation of two states, is calculated by employing the
channel-state duality. The interference term in the joint probability can arise
when the composite events form an entangled prospect and the system state is also
entangled. This interference is due to the uncertainty induced by nondestructive
measurements.

\end{abstract}

\vskip 1cm


\vskip 2mm


\vskip 2mm

Corresponding author: V.I. Yukalov
e-mail: yukalov@theor.jinr.ru

\newpage

\section{Introduction}

Multimode quantum systems provide efficient tools for quantum information processing
and quantum computing \cite{Nielsen_1,Keyl_2}. Multimode systems are ubiquitous. These
can be atomic systems with several populated electron energy levels, molecular ensembles
with several roto-vibrational modes, quantum dots with several exciton states, spin
assemblies with several spin projections, trapped Bose-condensed atomic gases with several
coherent modes, and so on \cite{Birman_3}. The general requirement for creating such
multimode systems is the existence of a discrete spectrum, which, in turn, usually
requires the finiteness of the quantum system. The generation of several modes can be
done by means of alternating resonance fields.

The probability of having a particular mode, at each moment of time, is characterized
by fractional mode populations. But, for the purpose of quantum information processing,
one may need to define the joint probability of observing two modes, or more generally,
of two states that can be entangled. The aim of the present paper is to introduce,
following the theory of quantum measurements for composite events \cite{Yukalov_3},
such a joint probability and to illustrate its properties by considering a two-mode
system of trapped atoms at low temperature with weak interactions, when almost all atoms
in an equilibrium trap would be in the Bose-Einstein condensate. Other modes can then be
generated by a resonance trap modulation.

In Sec. 2, we define the joint probability of two modes by employing the channel-state
duality based on the Choi-Jamiolkowski isomorphism. The general consideration is
specified in Sec. 3 for a multi-mode Bose-condensed system of trapped atoms. In
conclusion, we mention the analogies between the joint probabilities in the considered
multimode system and the processes studied in quantum decision theory and the problem
of creating quantum artificial intelligence.

\section{Multimode quantum systems}

A quantum system can be characterized by a set $\{|n\rangle\}$ of stationary solutions
corresponding to the eigenvectors of the related stationary Hamiltonian. Such stationary
solutions, associated with the eigenenergies $E_n$, are called modes, $n$ being a quantum
multi-index labeling the stationary states. The closed linear envelope of these modes
forms a Hilbert space
\be
\label{1}
 \cH = {\rm span} \{ | n \rgl \} \;  .
\ee
The system state is its statistical operator that, for zero temperature, can be expressed
as
\be
\label{2}
 \hat\rho(t)  = | \psi(t) \rgl \lgl \psi(t) |
\ee
through the wave functions
\be
\label{3}
 | \psi(t) \rgl = \sum_n c_n(t) | n \rgl \;  .
\ee
The coefficients $c_n$ are normalized, such that their moduli squared give the
fractional mode populations
\be
\label{4}
 f_n(t) \equiv | c_n(t)|^2 \; , \qquad \sum_n f_n(t) = 1\;  .
\ee

We may define as a simple event $A_n$ the observation of the mode $|n\rangle$ at
time $t$. According to the quantum theory of measurements \cite{Neumann_4}, this
event induces the correspondence
\be
\label{5}
A_n ~ \ra ~ | n \rgl ~ \ra ~ \hat P_n \equiv  | n \rgl \lgl n |  \; ,
\ee
where ${\hat P}_n$ is a projector. The probability of the event $A_n$ is
\be
\label{6}
 p(A_n) \equiv {\rm Tr}_\cH \hat\rho(t) \hat P_n = f_n(t) \;  ,
\ee
which, not surprisingly, coincides with the fractional mode population (4).

Let us now consider the case when more than a single measurement is done,
namely a series of measurements are accomplished at times
$0, t_1, t_2, \ldots, t_0, \ldots, t$. And suppose we wish to define the joint
probability of two events, one being the event $A_n$ of observing the $n$-mode
at time $t$, and another being an event $B_\alpha$ of observing a mode
$|\alpha\rangle$ at a preceding time $t_0 < t$. For the event $B_\alpha$, we
have the correspondence
\be
\label{7}
 B_\al ~ \ra ~ | \al \rgl ~ \ra ~ \hat P_\al \equiv | \al \rgl \lgl \al | \; .
\ee
Generally, the modes at different times could be of different types, if the
Hamiltonian has been changed. The events corresponding to the modes of different
types can be termed incompatible. But even if the events, hence the modes, are
compatible, that is, the modes are of the same type, it is convenient to
distinguish the modes for time $t_0$ from the modes associated with time $t$.

The definition of quantum joint probability requires caution. It is easy to see
that the Kirkwood \cite{Kirkwood_5} form does not constitute a probability, since
for incompatible events it can be complex valued, while for compatible modes it
reduces to the trivial expression
$$
{\rm Tr}_\cH \hat\rho(t) \hat P_n \hat P_\al = \dlt_{n\al} | c_n(t)|^2 \; .
$$

Even more complicated is the problem of defining the quantum joint probability
of two events, when one of them, say $B$, is not simple but consists of a union
$\{B_\alpha\}$ of several events $B_\alpha$, so that the event $B$ corresponds
to the observation of a multimode state
\be
\label{8}
 | B \rgl = \sum_\al b_\al | \al\rgl \; , \qquad
B \equiv \biguplus_\al B_\al \;  ,
\ee
where $b_\alpha$ depends on time $t_0$. Then how could the joint probability
of the events $A_n$ and $B$ be defined?

At this point, it is necessary to make some comments concerning notation. An event
$B$ representing a set $\{B_\alpha\}$ could be denoted as $\bigcup_\alpha B_\alpha$,
keeping in mind that this event induces the correspondence
$$
B ~ \ra ~ | B \rgl ~ \ra ~ \hat P_B \equiv | B \rgl \lgl B | \; .
$$
But, if one were to forget about the above correspondence, then employing the notation
$\bigcup_\alpha B_\alpha$ could lead to a confusion, if one would assume that
$\bigcup_\alpha B_\alpha$ corresponded to $\sum_\alpha \hat P_\alpha$. In order
to avoid such a confusion, we denote the complex event $B$ as is done in Eq. (8).

When one considers an explicitly prescribed dynamical picture, with the given
evolution law and exactly formulated measurement procedures, then the whole system
dynamics at times $0, t_1, t_2, \ldots, t_0, \ldots, t$ corresponds to the
convolution of channels with the given states:
\be
\label{9}
  \hat\rho_0 ~ \ra ~ \hat\rho_1 ~ \ra ~ \hat\rho_2 ~ \ra ~ \ldots ~ \ra ~
\hat\rho_{t_0} ~ \ra ~ \ldots ~ \ra \hat\rho_t \; .
\ee
Generally, the transformation of ${\hat \rho}_k$ to ${\hat \rho}_{k+1}$ is not unitary,
since it may involve nonunitary actions of a measuring device. But the preceding
states serve as initial conditions for the following states. Thus, if ${\hat \rho}_{t_0}$
is given, then it can be treated as an initial condition for ${\hat \rho}_t$, so that
${\hat \rho}_t = {\hat \rho}_t({\hat \rho}_{t_0})$. Then, the probability of two events
$A_n$ and $B_\alpha$ in the dynamical channel picture is
\be
\label{10}
 p_n(t) = {\rm Tr}_\cH \hat\rho_t(\hat\rho_{t_0}) \hat P_n \;  ,
\ee
corresponding to the channel
\be
\label{11}
 \{ \cH , \hat\rho_{t_0} \} ~ \ra ~ \{ \cH , \hat\rho_t(\hat\rho_{t_0}) \} \; .
\ee
Separating in the probability (10) the term $f_n(t)$ due to the system unitary evolution,
not perturbed by measurements, we have
\be
\label{12}
  p_n(t) = f_n(t) + q_n(t) \; .
\ee

The above channel picture, according to the Choi-Jamiolkowski isomorphism
\cite{Choi_6,Jamiolkowski_7} can be equivalently represented as a composite event
in a composite system constructed as follows. Let the event $A_n$ be the observation
of an $n$-mode at time $t$ and the event $B_\alpha$ be the observation of an
$\alpha$-mode at a preceding time $t_0 < t$. We can define the Hilbert spaces
\be
\label{13}
\cH_A \equiv {\rm span} \{ | n \rgl \} \; , \qquad
\cH_B \equiv {\rm span} \{ | \al \rgl \}
\ee
and introduce their tensor product
\be
\label{14}
 \cH_{AB} \equiv \cH_A \; \bigotimes \;
\cH_B =  {\rm span} \{ | n \al \rgl \equiv | n \rgl \otimes | \al \rgl \} \; .
\ee
The composite event of observing $A_n$ and $B_\alpha$ is denoted as
$A_n \bigotimes B_\alpha$ that induces the correspondence
\be
\label{15}
 A_n \; \bigotimes \; B_\al ~ \ra ~ \hat P_n \; \bigotimes \; \hat P_\al =
 | n \al \rgl \lgl n \al | \; .
\ee
The joint probability of these events is
\be
\label{16}
p( A_n \bigotimes B_\al) \equiv
{\rm Tr}_{AB} \hat\rho_{AB} \hat P_n \bigotimes \hat P_\al \;  ,
\ee
where the trace is over the space (14).

If the matrix elements of ${\hat \rho}_{AB}$ have the form
\be
\label{17}
\rho_{mn}^{\al\bt} \equiv \lgl m\al | \hat\rho_{AB} | n\bt \rgl =
c_{m\al} c_{n\bt}^* \; ,
\ee
then the probability (16) becomes
\be
\label{18}
 p( A_n \bigotimes B_\al) =
\lgl n\al | \hat\rho_{AB} | n\al \rgl = | c_{n\al} |^2 \;  .
\ee
For the correct normalization of the probability, such that
$$
p(A_n) = \sum_\al  p( A_n \bigotimes B_\al)  \; , \qquad
p(B_\al) = \sum_n  p( A_n \bigotimes B_\al)  \; ,
$$
\be
\label{19}
\sum_n  p( A_n) =  \sum_\al  p( B_\al) = 1\;  ,
\ee
the properties
$$
\sum_\al |c_{n\al}|^2 = | c_n |^2 \; , \qquad
\sum_n |c_{n\al}|^2 = | c_\al |^2 \; ,
$$
\be
\label{20}
 \sum_{n\al} |c_{n\al}|^2 = \sum_n | c_n|^2 =
\sum_\al |c_{\al}|^2 = 1 \; ,
\ee
are required.

The Choi-Jamiolkowski isomorphism establishes the channel-state duality, according
to which channel (11) is isomorphic to the composite system
$$
 \{ \cH_{AB} , \hat\rho_{AB} \} \;  ,
$$
with the composite state ${\hat \rho}_{AB}$. The joint probability of two simple
events is given by Eq. (18). It is then straightforward to introduce the conditional
quantum probability
$$
 p(A_n| B_\al) \equiv \frac{ p( A_n \bigotimes B_\al)}{p(B_\al)}
$$
resorting to the Bayesian rule.

The composite event $A_n \bigotimes B_\alpha$, being the tensor product of two
simple events, is called {\it factorized}. A more complicated structure arises
when one of the events is not simple but is a union of several events. Let us
consider the event $B = \biguplus_\alpha B_\alpha$ of having the multimode state
$|B\rangle$ at time $t_0$. Then the composite event
\be
\label{21}
\pi_n = A_n \; \bigotimes \; \biguplus_\al B_\al
\ee
represents an {\it entangled prospect} inducing the correspondence
\be
\label{22}
 \pi_n ~ \ra ~ | \pi_n \rgl \equiv
| n \rgl \bigotimes | B \rgl ~ \ra ~ \hat P(\pi_n) \;  ,
\ee
where
\be
\label{23}
\hat P(\pi_n) \equiv | \pi_n \rgl \lgl \pi_n |
\ee
is a prospect operator. The latter is required to satisfy the resolution of unity
\be
\label{24}
  \sum_n \hat P(\pi_n) = \hat 1_{AB} \; ,
\ee
with ${\hat 1}_{AB}$ being the identity operator on the space (14). By definition,
a prospect operator ${\hat P}(\pi_n)$ is positive, but it is not necessarily a
projector. The family $\{ {\hat P}(\pi_n) \}$ of positive operators, satisfying
resolution (24), forms a positive operator-valued measure \cite{Holevo_8,Holevo_9}.

The prospect probability is
\be
\label{25}
p(\pi_n) \equiv {\rm Tr}_{AB} \hat\rho_{AB} \hat P(\pi_n) \;   .
\ee
Separating here the diagonal and nondiagonal parts with respect to the indices
$\alpha$ and $\beta$, we obtain
\be
\label{26}
 p(\pi_n)  = f(\pi_n) + q(\pi_n)   \; ,
\ee
where
\be
\label{27}
 f(\pi_n) = \sum_\al | b_\al |^2 p(A_n\bigotimes B_\al) \; , \qquad
 q(\pi_n) = \sum_{\al\neq \bt} b_\al^* b_\bt c_{n\al} c^*_{n\bt} \; .
\ee

In view of resolution (24), the probability (25) is normalized:
\be
\label{28}
  \sum_n p(\pi_n) = 1 \; .
\ee
The term $f(\pi_n)$ describes the quasiclassical prospect probability that is
also normalized as
\be
\label{29}
  \sum_n f(\pi_n) = \sum_\al | b_\al |^2 p( B_\al) = 1 \; .
\ee
And the second term in the right-hand side of the probability (26) is the
{\it interference term} caused by the mode interference. Because of normalizations
(28) and (29), we have
\be
\label{30}
  \sum_n q(\pi_n) = 0 \; .
\ee

According to the channel-state duality, the prospect probabilities in the channel
picture and in the composite-state representation must coincide:
\be
\label{31}
 p(\pi_n) = p_n(t) \; .
\ee
Let us consider the case where the modes of the multimode state (8) are equally
weighted, that is, $|b_\alpha|^2 = const$, which, in view of normalizations (19)
and (29) gives $|b_\alpha|^2 = 1$. Then from Eq. (27) it follows that
\be
\label{32}
 f(\pi_n) = | c_n|^2 = f_n(t) \;  ,
\ee
as defined by expression (4). Therefore, for the interference term, we have
\be
\label{33}
 q(\pi_n) = q_n(t) \; ,
\ee
which allows us to calculate this term by employing the dynamical channel picture.

We may notice that the interference term disappears when the system state is not
entangled, having the form
$$
\hat\rho_{AB} = \sum_{n\al} | c_{n\al}|^2 | n\al \rgl \lgl n\al | \;   .
$$
Hence, state entanglement is a necessary condition for a nonzero interference
term. But this is not a sufficient condition. For instance, the maximally entangled
Bell state
$$
\hat\rho_{AB} = \frac{1}{M} \; \sum_{mn} | mm \rgl \lgl nn |
$$
yields $q(\pi_n) = 0$.

The Bell state is entangled and is also generating entanglement. The measure of entanglement
production is defined \cite{Yukalov_10} as
\be
\label{34}
 \ep(\hat\rho_{AB} ) \equiv
\log \; \frac{||\hat\rho_{AB} ||}{||\hat\rho_{A}|| \; ||\hat\rho_{B}|| }\;  ,
\ee
where
\be
\label{35}
 \hat\rho_{A} \equiv {\rm Tr}_B \hat\rho_{AB} \; , \qquad
\hat\rho_{B} \equiv {\rm Tr}_A \hat\rho_{AB} \; .
\ee
For the Bell state with $M$ modes, this gives $\varepsilon({\hat \rho}_{AB}) =\log M$.
However, the interference term is zero. Summarizing the above properties, we come to
the following conclusion.

\vskip 2mm

{\bf Proposition}. {\it For the interference term $q(\pi_n)$ to be nonzero, it is
necessary (but not sufficient) that the prospect $\pi_n$ be entangled and the state
${\hat \rho}_{AB}$ be entangled}.

\vskip 2mm

\section{Trapped Bose-condensed gas}

As a concrete illustration of the above theory, we consider the system of trapped
Bose atoms that weakly interact with each other, so that
\be
\label{36}
 \rho^{1/3} a_s \ll 1 \;  ,
\ee
where $\rho$ is the mean atomic density and $a_s$ is the scattering length. At low
temperature $T \ra 0$, almost all atoms pile down to the ground state, thus forming
a Bose-Einstein condensate corresponding to an atomic coherent state. The properties
of such atomic gases have been intensively discussed in several books
\cite{Lieb_11,Pitaevskii_12,Letokhov_13,Pethick_14} and review articles
\cite{Courteille_15,Andersen_16,Yukalov_17,Bongs_18,Yukalov_19,Morsch_20,Posazhennikova_21,
Yukalov_22,Proukakis_23,Yurovsky_24,Moseley_25,Bloch_26,Yukalov_27,Yukalov_28,Yukalov_29}.

The system is characterized by the coherent field $\eta({\bf r}, t)$ normalized to
the number of atoms
\be
\label{37}
  N = \int | \eta(\br,t) |^2 d\br \; .
\ee
The coherent field plays the role of the condensate wave function satisfying the
nonlinear Schr\"{o}dinger equation
\be
\label{38}
 i \; \frac{\prt}{\prt t} \; \eta(\br,t) = \hat H(\br,t) \eta(\br,t) \; ,
\ee
with the nonlinear Hamiltonian
\be
\label{39}
\hat H(\br,t) = -\; \frac {\nabla^2}{2m} +
U(\br,t) + \Phi_0 |\eta(\br,t)|^2 \;   ,
\ee
in which $\Phi_0 \equiv 4 \pi a_s/m$ is the effective interaction strength. Here and
in what follows, the Planck constant is set to one.

The external potential consists of two parts,
\be
\label{40}
 U(\br,t) = U(\br) + V(\br,t) \;  ,
\ee
of a stationary trapping potential $U(\bf r)$ and of an additional time-dependent
potential of trap modulation, which can be taken in the form
\be
\label{41}
V(\br,t) = V_1(\br)\cos(\om t) + V_2(\br) \sin(\om t) \;   .
\ee

It is convenient to make the replacement
\be
\label{42}
 \eta(\br,t) = \sqrt{N} \; \vp(\br,t) \;  ,
\ee
introducing the function that is normalized to one:
\be
\label{43}
 \int | \vp(\br,t) |^2 d\br = 1 \;  .
\ee
Then, Eq. (38) transforms into
\be
\label{44}
   i \; \frac{\prt}{\prt t} \; \vp(\br,t) =
[ \hat H_0(\br) + V(\br,t) ] \vp(\br,t) \;,
\ee
where the stationary nonlinear Hamiltonian is
\be
\label{45}
 \hat H_0(\br) =
 -\; \frac {\nabla^2}{2m} + U(\br) + N \Phi_0 |\vp(\br,t)|^2 \;   .
\ee

The coherent modes are defined \cite{Yukalov_30} as the solutions to the stationary
eigenvalue problem
\be
\label{46}
  \hat H_0(\br)  \vp_n(\br) = E_n  \vp_n(\br) \;  .
\ee
The solution of the time-dependent equation (44) can be represented as an expansion
over the coherent modes,
\be
\label{47}
  \vp(\br,t) = \sum_n c_n(t)  \vp_n(\br) e^{-iE_nt} \; .
\ee
Substituting this expansion into Eq. (44) yields the equations for the coefficient
functions $c_n(t)$. We assume that, at the initial time, practically the whole system
is in a coherent condensed state and only one more level with the energy $E_2$
could be slightly populated, so that the initial condition is
\be
\label{48}
 c_n(0) = c_1(0) \dlt_{n1} + c_2(0) \dlt_{n2}  .
\ee
The excited level with $E_2$ is selected by the use of the resonance alternating
field with the frequency
\be
\label{49}
\om = E_2 - E_1
\ee
that is in resonance with the corresponding transition frequency.

Under the initial condition (48) and the resonance condition (49), only two modes
are involved in the dynamics resulting in the equations
\be
\label{50}
  i \; \frac{dc_1}{dt} = \al_{12} | c_2|^2 c_1 + \frac{1}{2}\; \bt_{12} c_2 \; ,
\qquad
i \; \frac{dc_2}{dt} = \al_{21} | c_1|^2 c_2 + \frac{1}{2}\; \bt_{12}^* c_1 \; ,
\ee
in which the notations
$$
\al_{mn} \equiv
N \Phi_0 \int | \vp_m(\br)|^2 \left [ 2 | \vp_n(\br)|^2 - | \vp_m(\br)|^2
\right ] \; d\br \; ,
$$
$$
\bt_{mn} \equiv \int \vp_m^*(\br) [ V_1(\br) - i V_2(\br) ] \vp_n(\br)\; d\br
$$
are used. In what follows, we shall need the parameters
\be
\label{51}
 \al \equiv \frac{1}{2} ( \al_{12} + \al_{21} ) \;  , \qquad
\bt \equiv |\bt_{12}| \; , \qquad
\dlt \equiv \frac{1}{2} \; (\al_{12}-\al_{21} ) \;  .
\ee

Let us make the transformation
\be
\label{52}
 c_1 = \sqrt{\frac {1-s}{2} } \; e^{i\ep_1 t} \; , \qquad
 c_2 = \sqrt{\frac {1+s}{2} } \; e^{i\ep_2 t} \; ,
\ee
defining the population imbalance $s$ and the phase difference $x$ by the
corresponding expressions
\be
\label{53}
 s\equiv | c_2|^2 - |c_1|^2 \; , \qquad x \equiv \ep_1 -\ep_2 \;  .
\ee
Then, Eqs. (50) reduce to
\be
\label{54}
 \frac{ds}{dt} = - \bt\;\sqrt{1-s^2}\; \sin x \; , \qquad
\frac{dx}{dt} = \al s + \frac{\bt s}{\sqrt{1-s^2}} \; \cos x + \dlt \; .
\ee
In what follows, we keep in mind repulsive interactions, because of which
the parameter $\alpha$ is positive. The parameter $\delta$ is small and can be
omitted.

It is convenient to measure time in units of $1/\alpha$ and to introduce the
pumping parameter $b \equiv \beta/ \alpha$ characterizing the strength of the
modulation field as compared to that of the effective atomic interactions.
We assume that the observation of coherent modes is accomplished by means of
{\it nondestructive} measurements that influence only the phases of the modes,
so that a measurement leads to a random shift of the phase of each mode.
The evolution equations (54) become
\be
\label{55}
 \frac{ds}{dt} = - b\; \sqrt{1-s^2}\; \sin x \; , \qquad
dx=  s \left ( 1  + \frac{b}{\sqrt{1-s^2}} \; \cos x \right ) dt  +
\sgm dW_t \;   ,
\ee
where $W_t$ is the standard Wiener process and $\sigma$ is the
standard deviation related to the random noise.

Equations (55) without noise exhibit solutions of two different types, depending
on the value of the pumping parameter $b$. The dynamic transition between the
qualitatively different regimes of motion happens on the critical surface given by
the separatrix
\be
\label{56}
  2b_c \left ( 1 + \sqrt{1-s_0^2}\; \cos x_0 \right ) = s_0^2 \; ,
\ee
where the critical $b_c$ depends on the initial conditions $s_0 = s(0), x_0 = x(0)$.
In the {\it subcritical regime}, or {\it mode locked regime}, when $b < b_c$,
the modes experience weak Rabi oscillations close to their initial values and the
population imbalance never crosses the zero line, so that, if $s(0) < 0$ then
$$
-1 \leq s(t) < 0 \qquad ( b < b_c ) \;  .
$$
In the {\it supercritical regime}, or {\it mode unlocked regime}, when $b > b_c$,
the modes display Josephson oscillations, such that the population imbalance varies
in the whole range of its validity:
$$
-1 \leq s(t) \leq 1 \qquad ( b > b_c ) \;    .
$$
In the presence of random perturbations, transitions between these regimes can
happen, which we study numerically.

We solve Eqs. (55) for $s(t)$ and $x(t)$, which define the mode probabilities
\be
\label{57}
p_1(t) = \frac{1-s(t)}{2} \; , \qquad
p_2(t) = \frac{1+s(t)}{2} \;   .
\ee
The unperturbed fractional mode populations are given by
\be
\label{58}
 f_n(t) = \lim_{\sgm\ra 0} p_n(t) \;  .
\ee
The interference terms are defined by the expression
\be
\label{59}
 q_n(t) = p_n(t) - f_n(t) \;  .
\ee
We also study the behavior of the average interference terms
\be
\label{60}
 \overline q_n(t) \equiv \frac{1}{t} \int_0^t q_n(\tau)\; d\tau \; .
\ee

Because of the relations
$$
p_1(t) + p_2(t) = 1 \; , \qquad q_1(t) + q_2(t) = 0 \;   ,
$$
it is sufficient to show the behavior of only $p_1(t), q_1(t)$, and ${\bar q}_1(t)$.
The corresponding temporal evolution of these quantities are presented in
Figs. 1 and 2 for subcritical regime and in Figs. 3 and 4 for supercritical regime.
In all the cases, we take the initial conditions $s_0 = -0.9$ and $x_0 = 0$,
corresponding to the situation when, at $t = 0$, almost all atoms are Bose-condensed.
The probabilities, as well as interference terms, strongly fluctuate. The average
interference terms at the beginning fluctuate, tending to a constant at large time
$t \ra \infty$.

In order to find out the limit of the average interference terms, let us consider
the case of a small pumping parameter $b \ll 1$. Then, as follows from Eqs. (55),
the variable $s$ can be treated as slow and $x$ as fast. In such a case, we
can resort to the averaging techniques \cite{Bogolubov_31,Yukalov_31}.

Generally, by definition (59), we have
\be
\label{61}
 q_1(t) = \frac{1}{2} \; \left [ s^{(0)}(t) - s(t) \right ] \;  ,
\ee
where
\be
\label{62}
 s^{(0)}(t) \equiv \lim_{\sgm\ra 0} s(t) \;  .
\ee
When, there is no pumping, so that $b = 0$, then $s(t) = s_0$. To find the behavior
of $s(t)$ in the first order with respect to $b$, we need to get $x$ in the zero
order with respect to $b$. In the zero order in $b$, the second of equations (55)
simplifies into
\be
\label{63}
  dx = s_0 dt + \sgm d W_t \; .
\ee
Under the initial condition $x(0) = 0$, its solution reads
\be
\label{64}
  x = s_0 t + \sgm W_t \; .
\ee
Substituting this into the equation for the slow variable and averaging the
right-hand side over stochastic fluctuations, we have
\be
\label{65}
  \frac{ds}{dt} = - b \; \sqrt{1-s_0^2} \; \lgl\lgl \sin x \rgl\rgl \; ,
\ee
with the double angle brackets representing the averaging over these fluctuations.

Using the property
$$
 \lgl\lgl \; \exp ( i\sgm W_t ) \; \rgl\rgl = \exp \left \{ -\; \frac{\sgm^2}{2} \;
 \lgl\lgl W_t^2 \rgl\rgl   \right \} \; ,
$$
with $\lgl\lgl W_t^2 \rgl\rgl  = t$, we obtain
$$
\lgl\lgl \; \sin( s_0 t + \sgm W_t )\; \rgl\rgl  = \sin ( s_0 t) e^{-\sgm^2 t/2} \; .
$$
Consequently, Eq. (65) transforms into
\be
\label{66}
 \frac{ds}{dt} = - b\; \sqrt{1-s_0^2} \; \sin(s_0 t) e^{-\sgm^2 t/2} \; .
\ee
The solution to this equation, with the initial condition $s(0) = s_0$, reads as
\be
\label{67}
 s = s_0 + \frac{4b\sqrt{1-s_0^2}}{4s_0^2 + \sgm^4} \;
\left \{ \left [ s_0 \cos(s_0 t) + \frac{\sgm^2}{2} \; \sin (s_0 t) \right ]\;
e^{-\sgm^2 t/2} - s_0 \right \} \;  .
\ee
Respectively, solution (62) becomes
\be
\label{68}
 s^{(0)} = s_0 + \frac{b\sqrt{1-s_0^2}}{s_0} \; \left [ \cos(s_0 t) -1 \right ] \;  .
\ee

Averaging over time, we find
$$
\lim_{t\ra\infty} \; \frac{1}{t} \int_0^t s(\tau)\; d\tau =
s_0 \left ( 1 \; - \; \frac{4b\sqrt{1-s_0^2}}{4s_0^2+\sgm^4} \right ) \; ,
$$
\be
\label{69}
\lim_{t\ra\infty} \; \frac{1}{t} \int_0^t s^{(0)}(\tau)\; d\tau =
s_0 \left ( 1 \; - \; \frac{b\sqrt{1-s_0^2}}{s_0^2} \right ) \;    .
\ee
Hence, the limit of the time-averaged quantity (61) is
\be
\label{70}
 \lim_{t\ra\infty} \overline q_1(t) = - \;
\frac{b\sgm^4\sqrt{1-s_0^2}}{2s_0(4s_0^2+\sgm^4)} \;  .
\ee
Because of condition (30), we have ${\bar q}_2(\infty) = - {\bar q}_1(\infty)$.

When $b/\sigma \ra 0$, the limit ${\bar q}_1(\infty) = 0$, since, as has been
mentioned above, there is no dynamics in the system. In the opposite case,
when $\sigma/b \ra 0$, again ${\bar q}_1(\infty) = 0$, because the measurements
do not induce mode interference. In the intermediate situation, when neither $b$
nor $\sigma$ are zero, the interference term also is not zero. For example, for
$b = 0.25, s_0 = - 0.9$ and $\sigma = 0.5$, expression (70) yields
$\lim_{t\ra\infty} \overline q_1(t) = 0.001146$. In the limit of large-noise
amplitude $\sigma \ra +\infty$, we have
$\lim_{t\ra\infty} \overline q_1(t) = -\frac{b \sqrt{1-s_0^2}}{2s_0}$. For
$b = 0.25$ and $s_0 = - 0.9$, this gives $\overline q_1(\infty) = 0.06054$.

\section{Conclusion}

We have presented a general formulation of the quantum joint probability
for a multimode system, based on the Choi-Jamiolkowski isomorphism
between the channel picture and a composite system, via tensorial products
of Hilbert spaces. We have shown that, in order for the interference terms
to be present, it is essential that these states be entangled, which is achievable
by means of modern techniques \cite{Dobek_32,Zha_33}. As has been mentioned in the
Introduction, multimode states can be created in a variety of experimental setups.

We have applied this formalism to a two-mode system of trapped Bose-condensed atoms
under the influence of an oscillatory resonant trap modulation field. Assuming
non-destructive measurements, whose impact is only to scramble partially the phases,
we have obtained the dynamics of the population imbalance $s(t)$ between the two
modes and of the interference terms $q_1(t) = - q_2(t)$. In particular, we find
a non-zero long-term average of the interference terms, as long as the pumping
parameter $b$ and the standard deviation $\sigma$ of the random noise are non zero.

The population imbalance $s(t)$ allows us to find the probability
$p_1(t) = (1-s(t))/2$ of observing the first mode at time $t$, while being subject,
at previous times, to measuring perturbations modeled by random noise in
the evolution of the phase difference. Because of the random noise, the state at
time $t_0 < t$ is not exactly known, and is assumed to be a two-mode state with equal
weights. In that sense, finding $p_1(t)$ corresponds to a measurement under
uncertainty. On the other side, the certain measurement is associated with defining
the fractional mode population $f_1(t)$ calculated for the case without noise.
The difference $q_1(t) = p_1(t) - f_1(t)$ in this dynamical picture imitates the
behavior of the interference term $q(\pi_1)$, which is justified by the channel-state
duality.

We would like to mention that measurements under uncertainty are analogous to taking
decisions under uncertainty, as studied in decision making. In quantum decision theory
\cite{Yukalov_34,Yukalov_35,Yukalov_36,Yukalov_37}, taking decisions under uncertainty
is also accompanied by the appearance of interference terms corresponding to
deliberations in the evaluation of several admissible choices. The correspondence
suggests that the uncertainty existing in decision making may be interpreted as hidden
nondestructive measurements and processes occurring in the brain, which is continuously
exposed to exogenous influences. The existence of such interference has also to be taken
into account in quantum information processing and the problem of creating quantum
artificial intelligence \cite{Yukalov_38}. In quantum decision theory, the terms
$q(\pi_n)$ are interpreted as attraction factors, describing the attitude of a decision
maker to the available prospects. A negative attraction factor implies subconscious
repulsion of the decision maker to the considered prospect, while a positive attraction
factor is associated with the decision maker subconscious attraction to the prospect.

In that sense, treating the considered model as a cartoon of a functioning intelligence,
it is possible to give the following interpretation for the signs of term (70).
First, remembering that $s_0 < 0$, we see that ${\bar q}_1(\infty)$ is positive, which
increases the probability of the system to be in the ground state. In other words,
in the presence of the mode interference caused by measurements, the ground state is
more attractive for the system. On the contrary, the negative value ${\bar q}_2(\infty)$,
related to the excited upper mode, demonstrates a negative attitude of the system
to the population of the higher mode in the presence of random perturbations.

In the framework of quantum decision theory \cite{Yukalov_34,Yukalov_35,Yukalov_36,Yukalov_37},
it was shown that a non-informative prior leads to the {\it quarter law}, when
the average absolute value of the interference factor $q$ is equal to $1/4$.
Considering the present two-mode system of trapped Bose-condensed atoms under
the influence of an oscillating resonant trap modulation as a cartoon model of
functioning decision making, this would lead, in the limit of large noise amplitude
$\sigma \ra +\infty$, to the relationship $s_0^2 = 4b^2/(4b^2 + 1)$, for which
$\lim_{t\ra\infty} \overline q_1(t) = 1/4$. In other words, this suggests the existence
of a non trivial relationship between the initial state of the deciding mind and the
coupling with the external world. While being very tentative, this opens the road towards
a dynamical approach to quantum decision theory. Going beyond the non-informative prior,
a priori knowledge could be encoded into different initial states, leading to different
values of the attraction factors. These analogies suggest that cold trapped bosons in
the Bose-condensed state could be used for quantum information processing and for creating
quantum artificial intelligence.

\vskip 2mm

{\bf Acknowledgments}

\vskip 2mm

Financial support from the Swiss National Foundation and from the Russian Foundation
for Basic Research are appreciated.

\newpage

\newpage

\begin{center}
{\Large{\bf Figure Captions} }
\end{center}

\vskip 3cm

{\bf Fig.1} Subcritical regime with $b = 0.25 < b_c = 0.282$. Population
imbalance $s(t)$ and phase difference $x(t)$, as functions of
dimensionless time, for $\sgm = 0.5$ (solid line) and $\sgm = 0$ (dashed line).

\vskip 2cm
{\bf Fig.2} Subcritical regime with $b = 0.25 < b_c = 0.282$. Ground-state
probability $p_1(t)$ for $\sgm = 0.5$ (solid line) and $\sgm = 0$ (dashed line);
interference factor $q_1(t)$ and mean interference factor $\overline q_1(t)$
for $\sgm = 0.5$.

\vskip 2cm
{\bf Fig.3} Supercritical regime with $b = 0.5 > b_c = 0.282$. Population
imbalance $s(t)$ and phase difference $x(t)$, as functions of dimensionless
time, for $\sgm = 0.5$ (solid line) and $\sgm= 0$ (dashed line).

\vskip 2cm
{\bf Fig.4} Supercritical regime with $b = 0.5 > b_c = 0.282$. Ground-state
probability $p_1(t)$ for $\sgm = 0.5$ (solid line) and $\sgm = 0$ (dashed line);
interference factor $q_1(t)$ and mean interference factor $\overline q_1(t)$
for $\sgm = 0.5$.

\newpage

%Figure 1
\begin{figure}[ht]
\vspace{9pt}
\centerline{
\hbox{ \includegraphics[width=8cm]{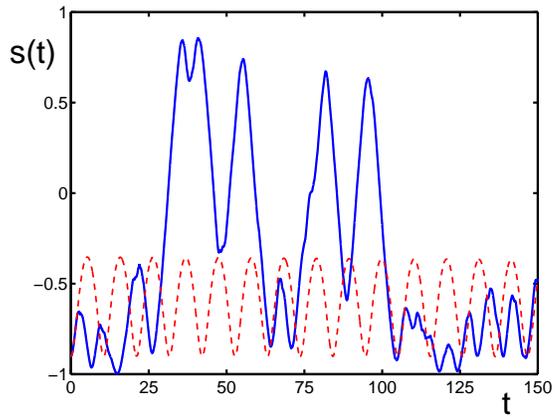} \hspace{2cm}
\includegraphics[width=8cm]{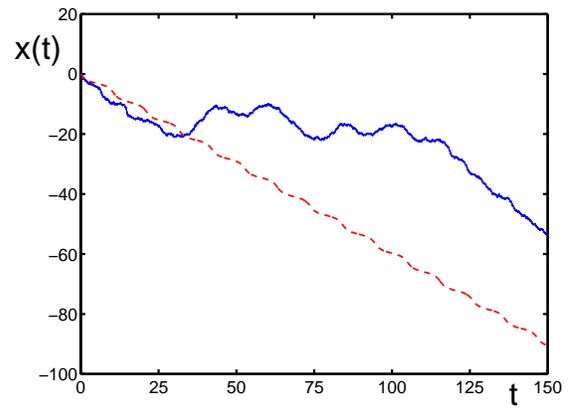} } }
\vspace{9pt}
\caption{Subcritical regime with $b = 0.25 < b_c = 0.282$. Population
imbalance $s(t)$ and phase difference $x(t)$, as functions of
dimensionless time, for $\sgm = 0.5$ (solid line) and $\sgm = 0$ (dashed line).
}
\label{fig:Fig.1}
\end{figure}

\newpage

%Figure 2
\begin{figure}[ht]
\vspace{9pt}
\centerline{
\hbox{ \includegraphics[width=8cm]{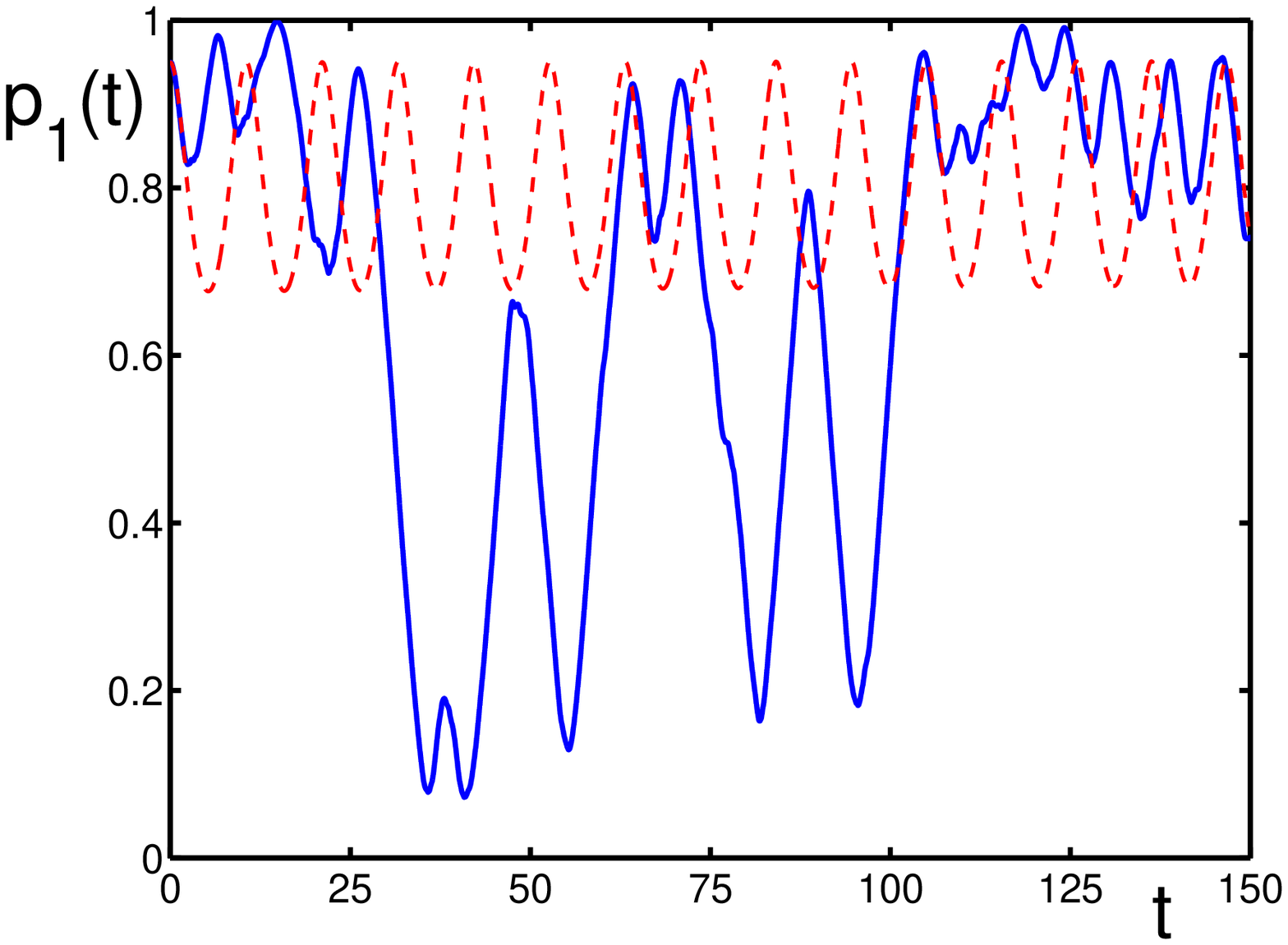} } }
\centerline{
\hbox{ \includegraphics[width=8cm]{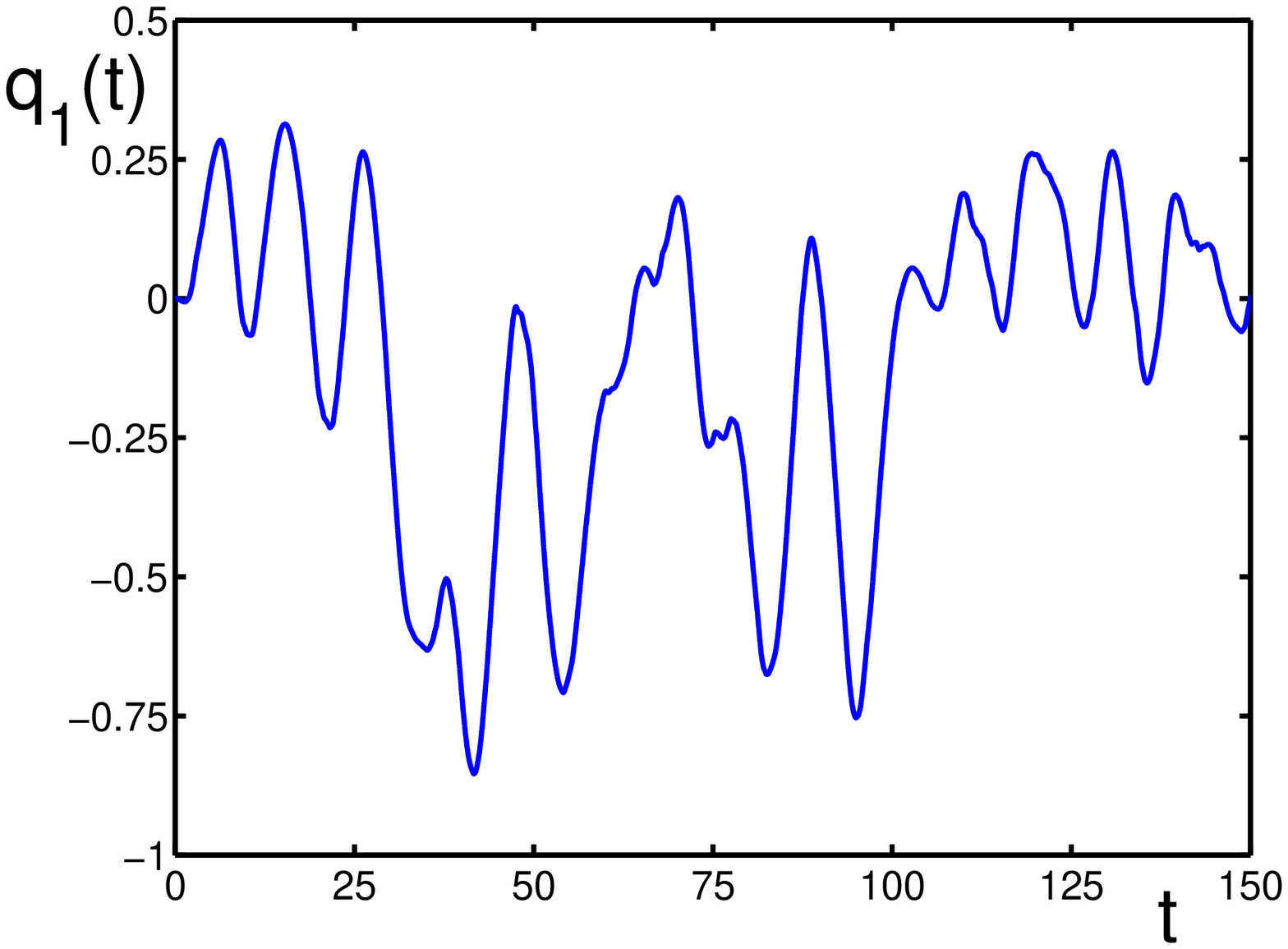} \hspace{2cm}
\includegraphics[width=8cm]{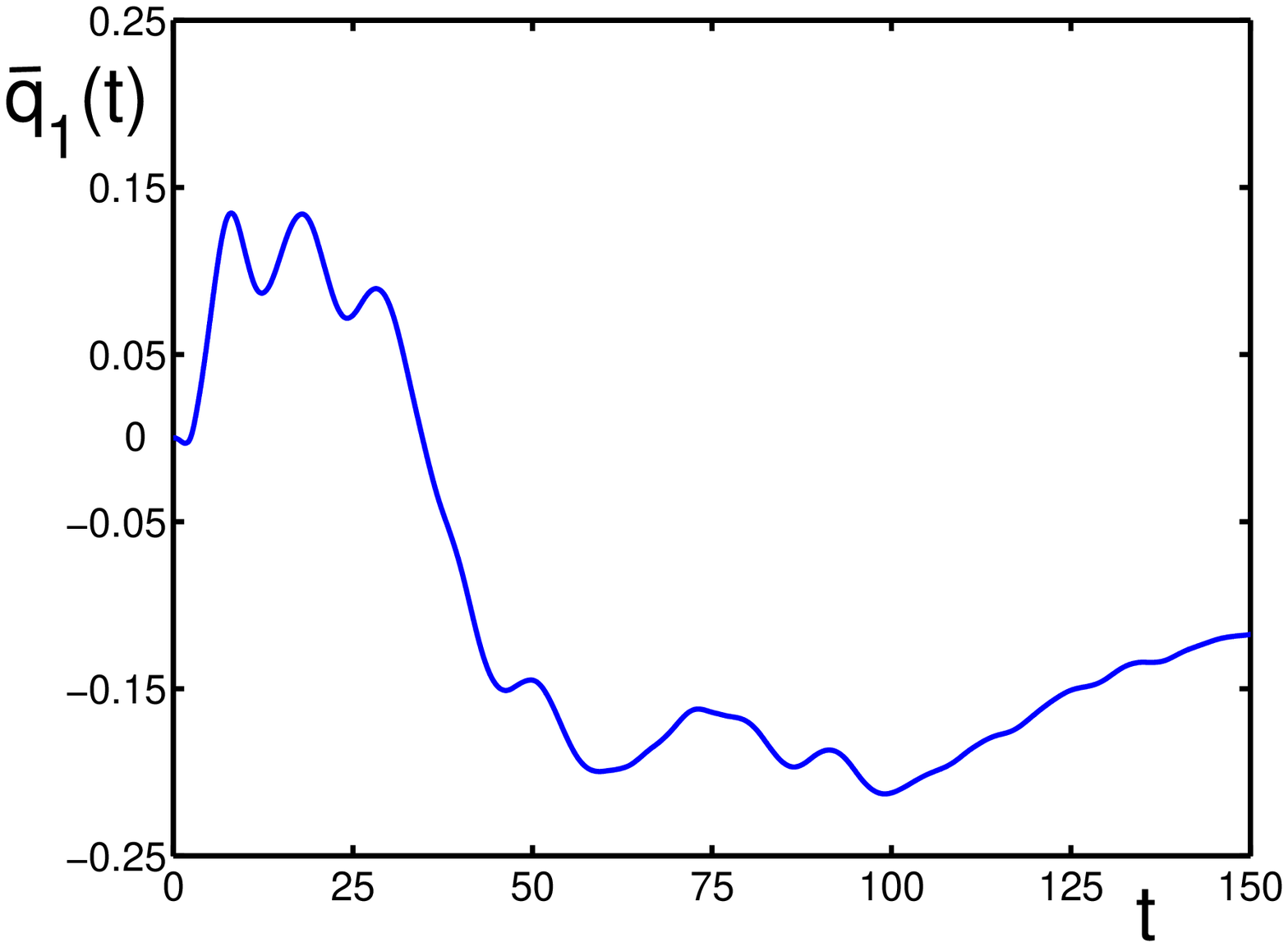} } }
\vspace{9pt}
\caption{Subcritical regime with $b = 0.25 < b_c = 0.282$. Ground-state
probability $p_1(t)$ for $\sgm = 0.5$ (solid line) and $\sgm = 0$ (dashed line);
interference factor $q_1(t)$ and mean interference factor $\overline q_1(t)$
for $\sgm = 0.5$.
}
\label{fig:Fig.2}
\end{figure}

\newpage

%Figure 3
\begin{figure}[ht]
\vspace{9pt}
\centerline{
\hbox{ \includegraphics[width=8cm]{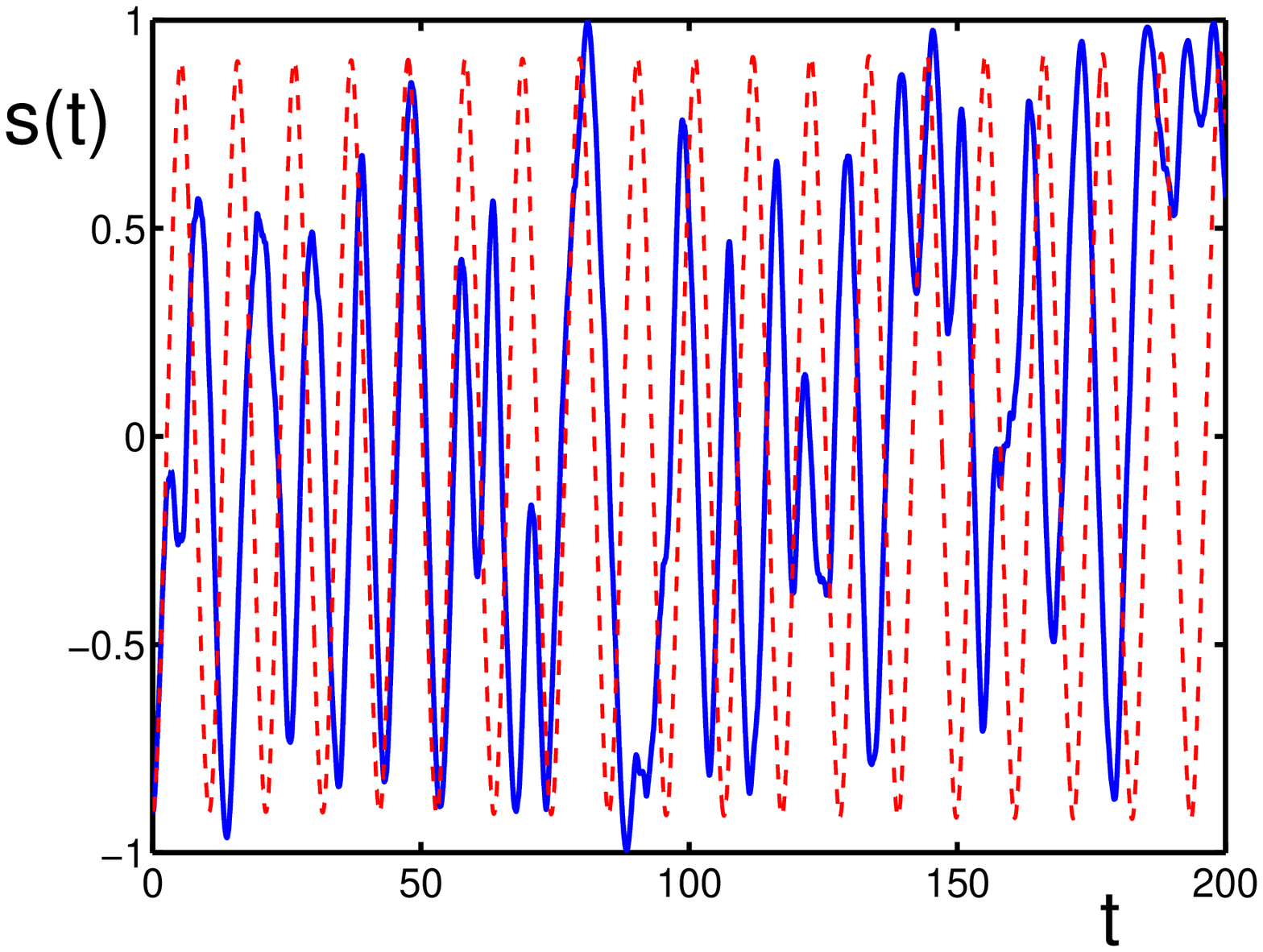} \hspace{2cm}
\includegraphics[width=7.5cm]{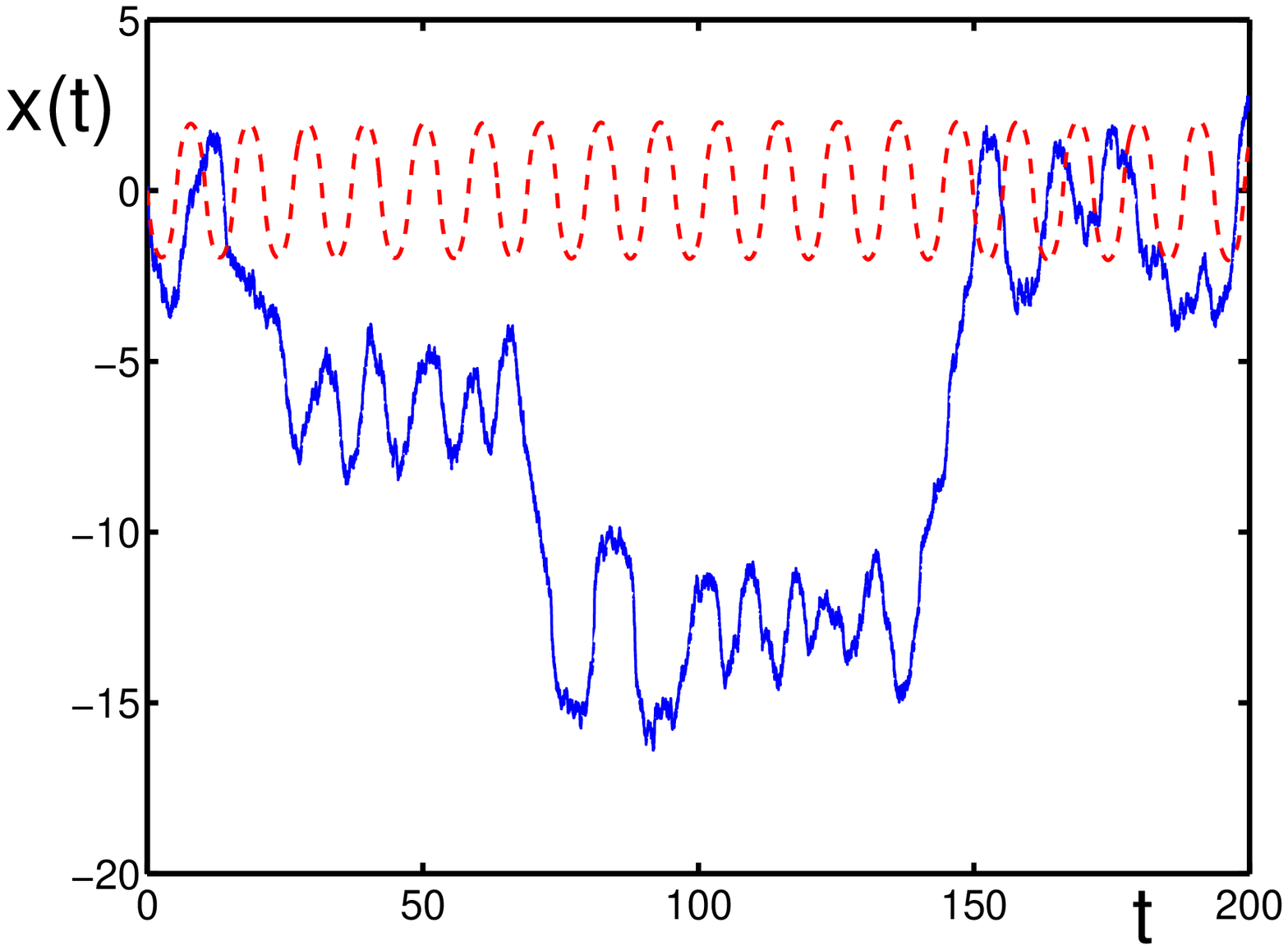} } }
\vspace{9pt}
\caption{Supercritical regime with $b = 0.5 > b_c = 0.282$. Population
imbalance $s(t)$ and phase difference $x(t)$, as functions of dimensionless
time, for $\sgm = 0.5$ (solid line) and $\sgm= 0$ (dashed line).
}
\label{fig:Fig.3}
\end{figure}

\newpage

%Figure 4
\begin{figure}[ht]
\vspace{9pt}
\centerline{
\hbox{ \includegraphics[width=8cm]{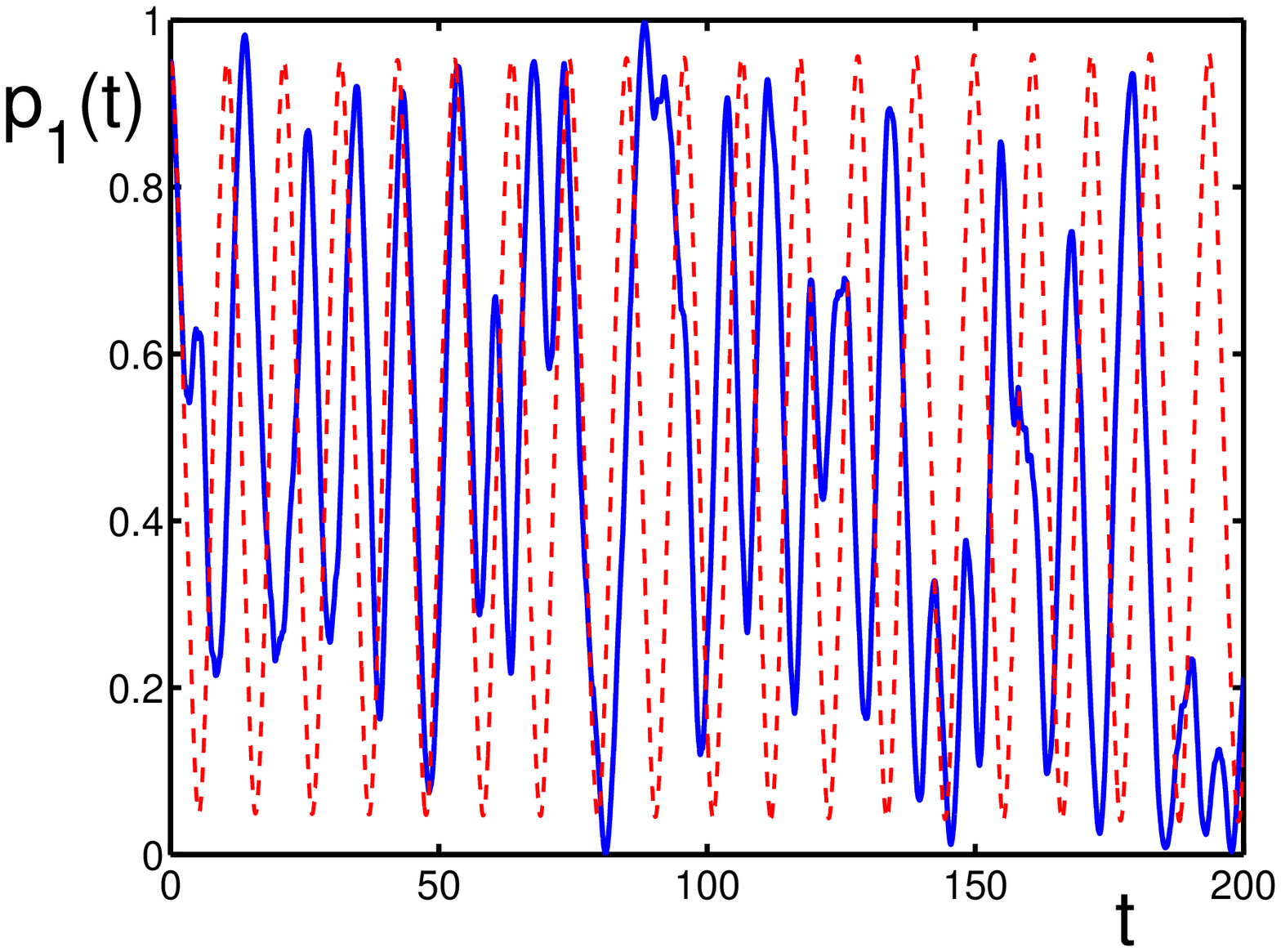} } }
\centerline{
\hbox{ \includegraphics[width=8cm]{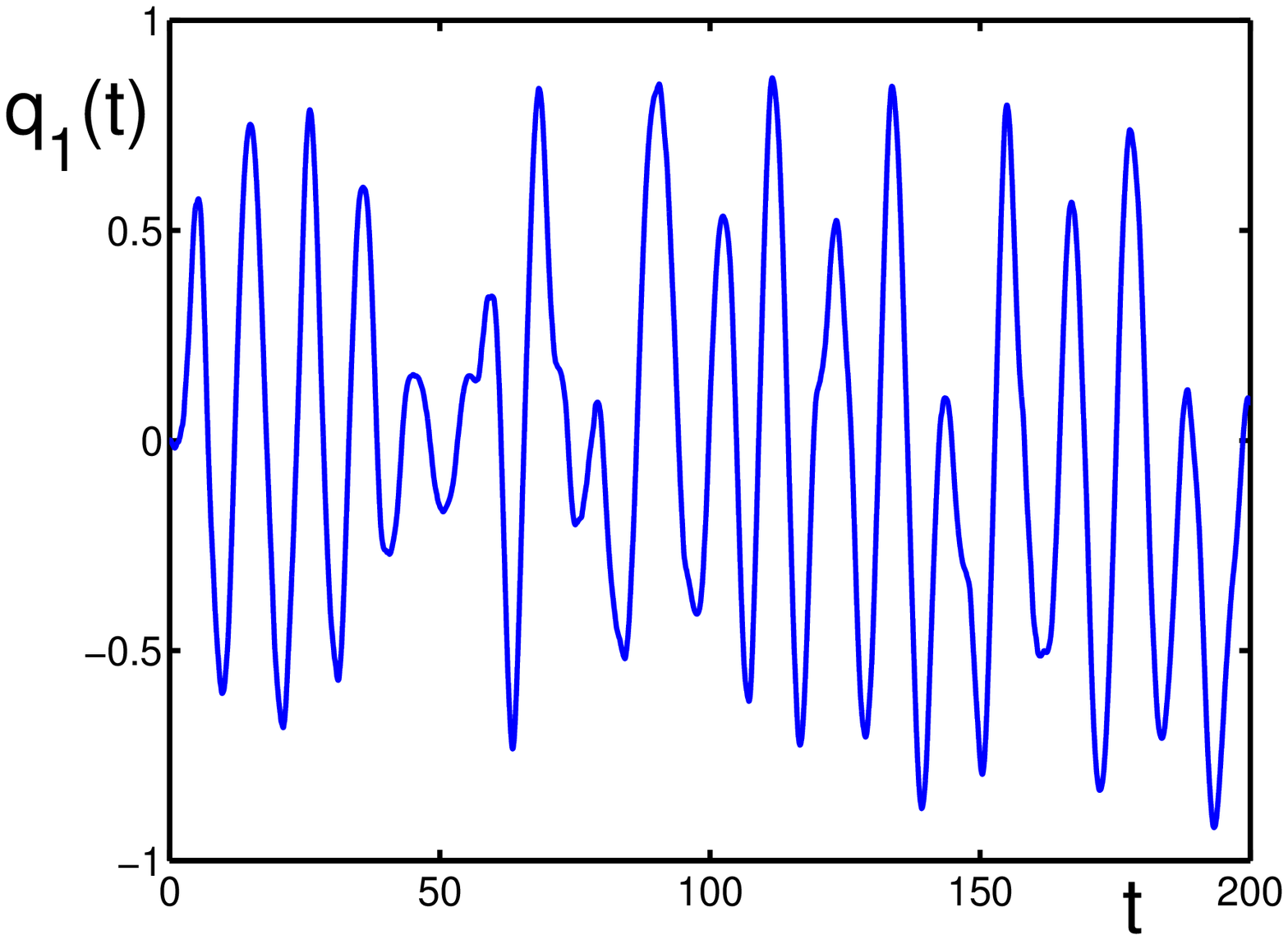} \hspace{2cm}
\includegraphics[width=7.5cm]{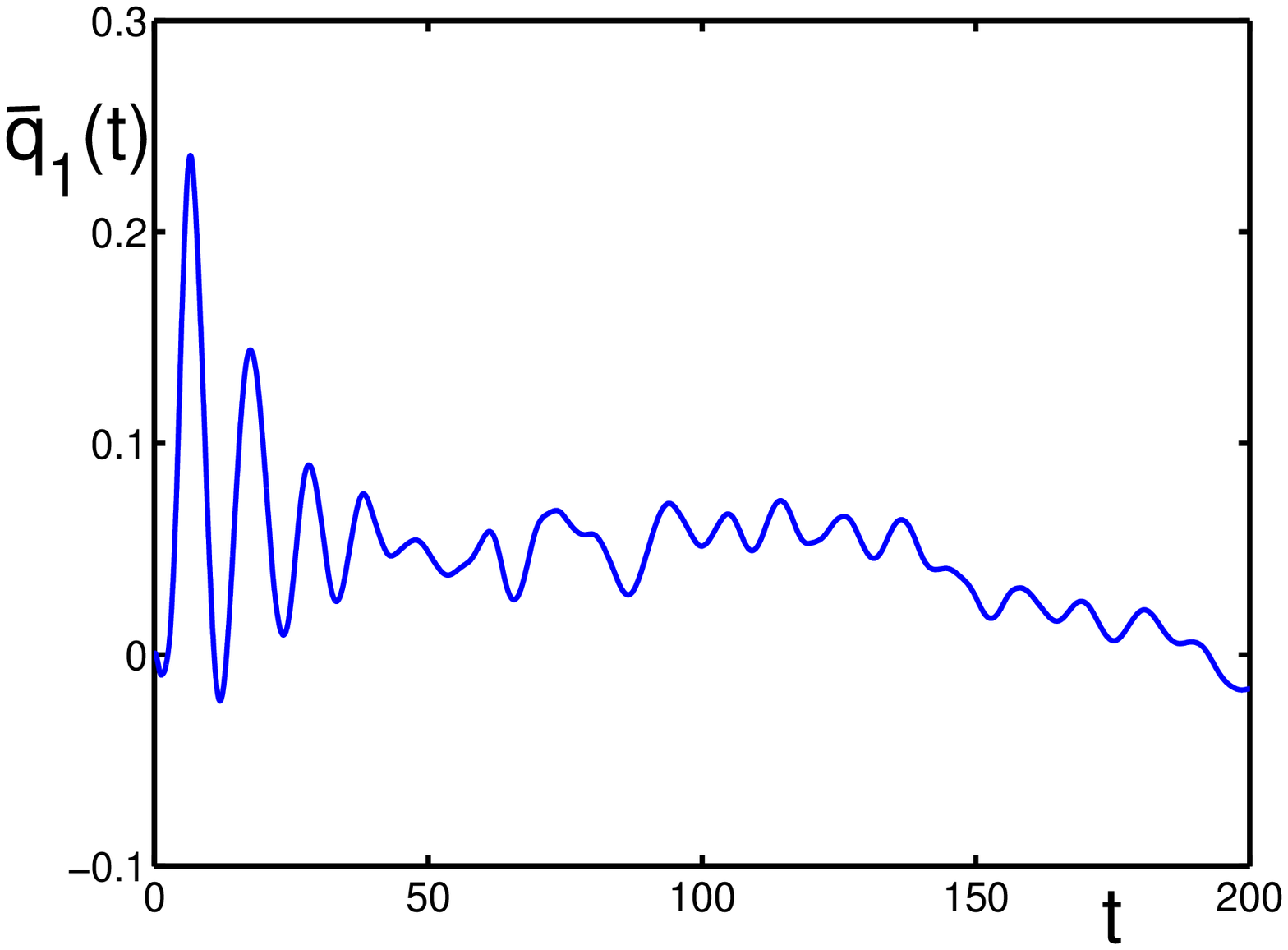} } }
\vspace{9pt}
\caption{Supercritical regime with $b = 0.5 > b_c = 0.282$. Ground-state
probability $p_1(t)$ for $\sgm = 0.5$ (solid line) and $\sgm = 0$ (dashed line);
interference factor $q_1(t)$ and mean interference factor $\overline q_1(t)$
for $\sgm = 0.5$.
}
\label{fig:Fig.4}
\end{figure}


\newpage


\begin{thebibliography}{99}


\bibitem{Nielsen_1}
Nielsen M and Chuang I 2000
{\it Quantum Computation and Quantum Information}
(Cambridge: Cambridge University)

\bibitem{Keyl_2}
Keyl M 2002
{\it Phys. Rep.} {\bf 369} 431

\bibitem{Birman_3}
Birman J L, Nazmitdinov R G and Yukalov V I 2013
{\it Phys. Rep.} {\bf 526} 1

\bibitem{Yukalov_3}
Yukalov V I and Sornette D 2013
{\it Laser Phys.} {\bf 23} 105502

\bibitem{Neumann_4}
von Neumann J 1955
{\it Mathematical Foundations of Quantum Mechanics}
(Princeton: Princeton University)

\bibitem{Kirkwood_5}
Kirkwood J G 1933
{\it Phys. Rev.} {\bf 44} 31

\bibitem{Choi_6}
Choi M D 1972
{\it Can. J. Math.} {\bf 24} 520

\bibitem{Jamiolkowski_7}
Jamiolkowski A 1972
{\it Rep. Math. Phys.} {\bf 3} 275

\bibitem{Holevo_8}
Holevo A S 2011
{\it Probabilistic and Statistical Aspects of Quantum Theory}
(Berlin: Springer)

\bibitem{Holevo_9}
Holevo A S and Giovannetti V 2012
{\it Rep. Prog. Phys.} {\bf 75} 046001

\bibitem{Yukalov_10}
Yukalov V I 2003
{\it Phys. Rev. A} {\bf 68} 022109

\bibitem{Lieb_11}
Lieb E H, Seiringer R, Solovej J P and Yngvason J 2005
{\it The Mathematics of the Bose Gas and its Condensation}
(Basel: Birkh\'{a}user)

\bibitem{Pitaevskii_12}
Pitaevskii L and Stringari S 2003
{\it Bose-Einstein Condensation} (Oxford: Clarendon)

\bibitem{Letokhov_13}
Letokhov V 2007
{\it Laser Control of Atoms and Molecules}
(New York: Oxford University Press)

\bibitem{Pethick_14}
Pethick C J and Smith H 2008
{\it Bose-Einstein Condensation in Dilute Gases}
(Cambridge: Cambridge University Press)

\bibitem{Courteille_15}
Courteille P W, Bagnato V S and Yukalov V I 2001
{\it Laser Phys.} {\bf 11} 659

\bibitem{Andersen_16}
Andersen J O 2004
{\it Rev. Mod. Phys.} {\bf 76} 599

\bibitem{Yukalov_17}
Yukalov V I 2004
{\it Laser Phys. Lett.} {\bf 1} 435

\bibitem{Bongs_18}
Bongs K and Sengstock K 2004
{\it Rep. Prog. Phys.} {\bf 67} 907

\bibitem{Yukalov_19}
Yukalov V I and Girardeau M D 2005
{\it Laser Phys. Lett.} {\bf 2} 375

\bibitem{Morsch_20}
Morsch O and Oberthaler M 2006
{\it Rev. Mod. Phys.} {\bf 78} 179

\bibitem{Posazhennikova_21}
Posazhennikova A 2006
{\it Rev. Mod. Phys.} {\bf 78} 1111

\bibitem{Yukalov_22}
Yukalov V I 2007
{\it Laser Phys. Lett.} {\bf 4} 632

\bibitem{Proukakis_23}
Proukakis N P and Jackson B 2008
{\it J. Phys. B: At. Mol. Opt. Phys.} {\bf 41} 203002

\bibitem{Yurovsky_24}
Yurovsky V A, Olshanii M and Weiss D S 2008
{\it Adv. At. Mol. Opt. Phys.} {\bf 55} 61

\bibitem{Moseley_25}
Moseley C, Fialko O and Ziegler K 2008
{\it Ann. Phys. (Berlin)} {\bf 17} 561

\bibitem{Bloch_26}
Bloch I, Dalibard J and Zwerger W 2008
{\it Rev. Mod. Phys.} {\bf 80} 885

\bibitem{Yukalov_27}
Yukalov V I 2009
{\it Laser Phys.} {\bf 19} 1

\bibitem{Yukalov_28}
Yukalov V I 2011
{\it Phys. Part. Nucl.} {\bf 42} 460

\bibitem{Yukalov_29}
Yukalov V I 2012
{\it Laser Phys.} {\bf 22} 1145

\bibitem{Yukalov_30}
Yukalov V I, Yukalova E P and Bagnato V S 1997
{\it Phys. Rev. A} {\bf 56} 4845

\bibitem{Bogolubov_31}
Bogolubov N N and Mitropolsky Y A 1961
{\it Asymptotic Methods in the Theory of Nonlinear Oscillations}
(New York: Gordon and Breach)

\bibitem{Yukalov_31}
Yukalov V I and Yukalova E P 2000
{\it Phys. Part. Nucl.} {\bf 31} 561

\bibitem{Dobek_32}
Dobek K, Karpinski M, Demkowicz-Dobrzanski R, Banaszek K and Horodecki P 2013
{\it Laser Phys.} {\bf 23} 025204

\bibitem{Zha_33}
Zha X, Yuan C and Zhang Y 2013
{\it Laser Phys. Lett.} {\bf 10} 045201

\bibitem{Yukalov_34}
Yukalov V I and Sornette D 2008
{\it Phys. Lett. A} {\bf 372} 6867

\bibitem{Yukalov_35}
Yukalov V I and Sornette D 2009
{\it Eur. Phys. J. B} {\bf 71} 533

\bibitem{Yukalov_36}
Yukalov V I and Sornette D 2009
{\it Entropy} {\bf 11} 1073

\bibitem{Yukalov_37}
Yukalov V I and Sornette D 2010
{\it Adv. Complex Syst.} {\bf 13} 659

\bibitem{Yukalov_38}
Yukalov V I and Sornette D 2009
{\it Laser Phys. Lett.} {\bf 6} 833

\end{thebibliography}
\end{document}